\documentclass[10pt,oneside]{amsart}
\usepackage{amstext,amsmath,amsfonts,amscd,amsthm}
\usepackage{cite}

\begin{document}
\title{Thermal Field Dynamics and Bialgebras}

\author[T. Kopf]{T. Kopf}
\address{Tom\'a\v{s} Kopf, Theoretical Physics Institute,  Avadh Bhatia Phys. Lab., University of Alberta, Edmonton, Alberta T6G 2J1, Canada}

\author[A. E. Santana]{A. E. Santana}
\address{Ademir E. Santana, Theoretical Physics Institute,  Avadh Bhatia Phys. Lab., University of Alberta, Edmonton, Alberta T6G 2J1, Canada\\
and Instituto de Fisica, Universidade Federal da Bahia, Campus de Ondina, 40210-340, Salvador, Bahia, Brazil}

\author[F. C. Khanna]{F. C. Khanna}
\address{Faqir C. Khanna, Theoretical Physics Institute,  Avadh Bhatia Phys. Lab., University of Alberta, Edmonton, Alberta T6G 2J1, Canada\\
and TRIUMF, 4004 Wesbrook Mall, Vancouver, British Columbia, Canada, V6T 2A3}

\begin{abstract}
In Thermal Field Dynamics, thermal states are obtained from
restrictions of vacuum states on a doubled field algebra. It is shown
that the suitably doubled Fock representations of the Heisenberg
algebra do not need to be introduced by hand but can be canonically
handed down from deformations of the extended Heisenberg bialgebra. No
artificial redefinitions of fields are necessary to obtain the
thermal representations and the case of
arbitrary dimension is considered from the beginning. Our results
support a possibly fundamental role of bialgebra structures in
defining a general framework for Thermal Field Dynamics.
\end{abstract}

\maketitle

\section{Introduction}

Thermal Field Dynamics \cite{Umezawa} is based on the idea that thermal states of a quantum system, described by the field algebra $\mathbf{A}$ can be
given as restrictions of vacuum states of a doubled algebra of
observables $\mathbf{A\otimes A}$. The doubling of $\mathbf{A}$ is usually
given by the so called tilde conjugation rules which can be thought of as a 
mapping of ${\mathbf{A}} \cong {\mathbf{A}}\otimes \mathbf{1}$ into
$\mathbf{1}\otimes{\mathbf{A}}$:

\begin{align*}
{(ab)}^{\sim} &= \tilde{a}\tilde{b}\\
{(\lambda a + \mu b)}^{\sim} &= {\lambda}^{\ast} \tilde{a} + {\mu}^{\ast}
\tilde{b}\\
{(\tilde{a})}^{\sim} &= a \\
{({a}^{+})}^{\sim} &= {\tilde{a}}^{+}\\
{\mid vacuum \rangle}^{\sim} &= \mid vacuum \rangle\\
&\text{with $a,b \in \mathbb{C}$ and $\lambda ,\mu \in {\mathbf{A}}$}
\end{align*}

However the explicit splitting of ${\mathbf{A\otimes A}}$ into ${\mathbf{A}}\otimes
\mathbf{1}$ and $\mathbf{1}\otimes\mathbf{A}$ is not necessary. For all
intended purposes a doubling from $\mathbf{A}$ to $\mathbf{A\otimes A}$ without
a specification of two copies of $\mathbf{A}$ in $\mathbf{A\otimes A}$ is
sufficient and one can therefore in all the following drop the tilde conjugation
rules retaining only the idea of a doubling.

Note, that already at this point the present treatment departs from a different and differently motivated approach to Thermal Field Dynamics based on the modular (To\-mi\-ta\--Ta\-ke\-sa\-ki) conjugation as given by I.Ojima \cite{Ojima}. There the modular conjugation realizes an omnipresent tilde conjugation by switching the algebra $\mathbf{A}$ and its commutant ${\mathbf{A}}^{\prime}$ in the thermal representation.

 Given a vacuum state ${\omega}_{J}$ on $\mathbf{A}$ one can introduce
a doubling
${\delta}_{\chi} :{\mathbf{A}}\rightarrow {\mathbf{A}}  \otimes {\mathbf{A}}$ such
that the desired thermal state ${\omega}_{\chi}$ is given by

\begin{equation}
{\omega}_{\chi} = ({\omega}_{J}\otimes
{\omega}_{J})\circ {\delta}_{\chi}
\end{equation}

That this is indeed the case is shown in section \ref{2} which reviews also basic
facts on the Heisenberg algebra and its exponentiation, the Weyl algebra,
mostly omitting proofs. This scheme of producing thermal states has in this
setting no deeper justification except that it works.

However, there are two more general points of view:

First, the doubling of the algebra of observables characteristic for Thermal
Field Dynamics is just an example of taking tensor products of representations
of the field algebra $\mathbf{A}$. A general and systematic way of taking
tensor products of representations is given, if $\mathbf{A}$ is equipped with a
bialgebra structure \cite{Madore}, so it would be nice to have one on our algebra.

Second, a bialgebra was argued to be possibly the right structure for an
axiomatic approach to Thermal Field Dynamics and may thus be assumed from the
beginning for a canonical construction of a  doubling of the field algebra
\cite{Khanna_Santana}. In that approach a part of the structure of the theory
is deduced from a bialgebra (eventually produced from a symmetry Lie algebra \cite{Khanna_Santana,Song-Ding-An})
with the rest given by the requirement of a Fock structure. The Fock structure
itself can, however, also be cast into a bialgebra form by using the extended Heisenberg algebra (see section \ref{3}) thus
allowing an axiomatic setting of the theory starting only from a bialgebra.

Both of these points of view ask for a bialgebra. Unfortunately, the Heisenberg
algebra $\mathbf{A}$ cannot be turned into one as shown in section \ref{3}. However
it turns out that by going over to a slightly different algebra, the extended
Heisenberg algebra $\mathbf{U}$, one can do away with this problem, and there is
a mapping from the extended Heisenberg algebra $\mathbf{U}$ onto the Heisenberg
algebra $\mathbf{A}$ that allows to transport interesting structures,
particularly the comultiplication $\Delta$ responsible for tensor products of
representations into the context of physical observables.

This is true even if one q-deforms the extended Heisenberg bialgebra: The
deformed extended Heisenberg algebra ${\mathbf{U}}_{\chi}$ maps down onto the
{\it undeformed} Heisenberg algebra $\mathbf{A}$ giving thus nothing new for
the algebra but providing us with new possible doublings ${\Delta}_{\chi}$. It
is these new doublings ${\Delta}_{\chi}$ coming from deformations of the
extended Heisenberg algebra that allow us to express an arbitrary quasifree
state in a way similar to Thermal Field Dynamics
{}.
This is shown in section \ref{3} together with a discussion of the relevant
deformations.
The fact that one can produce squeezed states and thermal states by deformations of the extended Weyl algebra has been shown using 1-dimensional examples   \cite{Iorio-Vitiello,Celeghini1,Celeghini2}. In particular, a coherent state representation has been exploited and  applications to lattice quantum mechanics have been suggested \cite{Celeghini1,Celeghini2}.

An example showing the relationship between the deformation parameter $\chi$
and the inverse temperature of the corresponding thermal state is given in
section \ref{4}. Section \ref{5} contains conclusions and some general remarks.

\section{The Heisenberg algebra and its Weyl form}\label{2}

In order to set the notation it will be shown now that thermal states
on a Heisenberg algebra $\mathbf{A}$  can be obtained from a Fock
representation  (i.e.  an  irreducible  representation obtainable
from     a     vacuum      ${\omega}_{J}$     by     the
Ge\'{l}fand-Naimark-Segal-construction   (GNS-construction,   see
e.g. in O.Bratteli and D.W.Robinson \cite{Bratteli-Robinson1}) by a suitable ad hoc doubling.
The Heisenberg algebra is generated from a
symplectic vector space $\Gamma$ of arbitrary dimension with the
symplectic form $\sigma (\bullet ,\bullet )$  by the usual commutation
relations:

\begin{align*}
\phi({z}_{1}) \phi({z}_{2}) - \phi({z}_{2}) \phi({z}_{1}) &= i\hbar
\sigma ({z}_{1},{z}_{2}) &{z}_{1}, {z}_{2} \in \Gamma \\
\intertext{or}
W({z}_{1})  W({z}_{2}) &= {e}^{{\frac i 2} \sigma ({z}_{1},{z}_{2})}
W({z}_{1}+{z}_{2}) &{z}_{1}, {z}_{2} \in \Gamma
\end{align*}

Here $\phi(z)$ are the field operators and $W(z)$ their exponentiated
Weyl form:

\begin{equation*}
W(z) ={e}^{i \phi(z)}
\end{equation*}

A vacuum state ${\omega}_{J}$ on the field algebra $\mathbf{A}$
is given by a complex structure $J$  on $\Gamma$. On the
Weyl generators one has:

\begin{equation}\label{vacuum}
{\omega}_{J}(W(z)) = {e}^{-{\frac 1 4} z\circ J\circ
\sigma\circ z}
\end{equation}

It is also fully determined by its two-point function:

\begin{equation*}
{\omega}_{J}(\phi({z}_{1}) \phi({z}_{2})) = {\frac 1 2}
{z}_{1}\circ J\circ \sigma \circ {z}_{2}
\end{equation*}

The thermal states to be considered are quasifree and thus
correspond to free (i.e. quadratic) field Hamiltonians. For each
quasifree state  there is by  the modular theory  \cite{Takesaki}
(Kubo-Martin-Schwinger-theory  (KMS-theory, see O.Bratteli and D.W.Robinson
\cite{Bratteli-Robinson1,Bratteli-Robinson2})) a
Hamiltonian with respect to which the state is thermal and therefore
one has to show that one can produce any quasifree state of interest by
our doubled Fock representations. Any quasifree state corresponding to
a positive definite Hamiltonian can be written in the form (compare with
O.Bratteli and D.W.Robinson \cite{Bratteli-Robinson2},p.50):

\begin{equation}
{\omega}_{\chi}(W(z)) = {e}^{-{\frac 1 4} z\circ
\coth{{\frac\Omega 2}} \circ J\circ\sigma\circ z}
\end{equation}

where $\Omega$ is a positive definite operator arising from the
diagonalization of the corresponding Hamiltonian $H$ \cite{Krtous}:

\begin{equation}\label{Hamiltonian}
H = \Omega\circ J \circ\sigma
\end{equation}

It is known from Thermal Field Dynamics that $\Omega$ can be related to a
Bogoljubov operator $\chi$ on $\Gamma$ anticommuting with $J$ by the
following relation,
giving an alternative parametrization of quasifree states:

\begin{align}
\cosh{(2 \chi)} &= \coth{{\frac\Omega 2}}\\
{\omega}_{\chi}(W(z)) &= {e}^{-{\frac 1 4} z\circ
\cosh{(2 \chi)} \circ J\circ\sigma\circ z}\label{quasifree}\\
\intertext{with $\chi$ satisfying:}
J\circ \chi &= - \chi \circ J \label{anticom1}\\
\sigma \circ \chi &= - \chi \circ \sigma \label{anticom2}
\end{align}

The promised doubling ${\delta}_{\chi}$ is now given by the
following action on the Weyl generators:

\begin{equation}
{\delta}_{\chi}(W(z)) = W(\cosh{(\chi)} z)\otimes W(\sinh{(\chi)} z)
\end{equation}

One can check now by direct calculation that ${\delta}_{\chi}$ is an
algebra homomorphism and that it produces from the doubled vacuum
state ${\omega}_{J}  \otimes {\omega}_{J}$  the right
quasifree state ${\omega}_{\chi}$:

\begin{align*}
{\omega}_{\chi} &= ({\omega}_{J}\otimes {\omega}_{J}) \circ
{\delta}_{\chi} (W(z))=&\\
&=({\omega}_{J}\otimes {\omega}_{J})(W(\cosh{\chi} z)\otimes
W(\sinh{\chi} z)) =&\\
&={\omega}_{J}(W(\cosh{\chi} z)) {\omega}_{J}(W(\sinh{\chi}
z)) =&\\
\intertext{using (\ref{vacuum})}
&={e}^{-{\frac 1 4} z\circ\cosh{\chi}\circ J\circ
\sigma\circ\cosh{\chi}\circ z}{e}^{-{\frac 1 4} z\circ\sinh{\chi}\circ
J\circ\sigma\circ\sinh{\chi}\circ z} =&\\
\intertext{by (\ref{anticom1}, \ref{anticom2})}
&={e}^{-{\frac 1 4} z\circ({\cosh{}}^{2}\chi +{\sinh{}}^{2}\chi )\circ
{J}\circ\sigma\circ z} =& \\
\intertext{by the identity ${\cosh{}}^{2}\chi +{\sinh{}}^{2}\chi = \cosh{2\chi}$}
&={e}^{-{\frac 1 4} z\circ\cosh{2\chi}\circ {J}\circ\sigma\circ z}&
\end{align*}

But this is just the quasifree state (\ref{quasifree}) that is required.

\section{The extended Heisenberg algebra and its deformations}\label{3}

One would like to use a bialgebra structure on a field algebra, in particular
the Heisenberg algebra $\mathbf{A}$, and by the  GNS-construction a vacuum state ${\omega}_{{J}}$
giving a Fock representation, to produce a new
representation in which the vacuum doubled by the comultiplication $\Delta$
will be a thermal (and thus reducible) state.

There is however a  problem with this straightforward idea: There is no
bialgebra structure on the Heisenberg algebra ${\mathbf A}$. This can be easily
seen from the fact that the Heisenberg commutation relations for the field
$\phi (z)$ require a commutator to be proportional to the unit of the algebra.

\begin{equation*}
\phi ({z}_{1})\phi ({z}_{2}) - \phi ({z}_{2})\phi ({z}_{1}) = i \hbar \sigma
({z}_{1},{z}_{2}) \mathbf{1}
\end{equation*}

 Now, a counit $ \varepsilon$ of the bialgebra structure has to vanish on
commutators and has to be equal to $1$ on the unit of the algebra which is not
possible unless the proportionality constant in the commutation relations
(Planck's constant) is zero.

To improve that, the unit $\mathbf{1}$ of the algebra can be replaced by an
abstract central element $H$. Now it is no more necessary for the counit $
\varepsilon$ to be equal to $\mathbf{1}$ on this central element, the
commutation relations can be considered as a Lie algebra and there exists even
a Hopf algebra structure on this extended Heisenberg algebra $\mathbf{U}$ which
is actually now a universal enveloping algebra of a Lie algebra \cite{Majid}.
To recover a meaning in the field algebra one
can map the extension $\mathbf{U}$ onto the plain Heisenberg algebra ${\mathbf
A}$. The map ${\Delta}^{()}:{\mathbf A}\rightarrow{\mathbf A}\otimes{\mathbf
A}$ induced from the comultiplication
$\Delta:\mathbf{U}\rightarrow\mathbf{U}\otimes\mathbf{U}$ is no more preserving
the algebra unit but it is a morphism of algebras and thus allows a 
tensor product of representations.

What is gained by considering  the extended Heisenberg algebra
$\mathbf{U}$ is the possibility of having an underlying bialgebra structure
giving a canonical doubling on the algebra of observables ${\mathbf A}$.

But moreover  the extended  Heisenberg algebra
$\mathbf{U}$ can be deformed without  changing the scheme and  thus producing new
interesting doublings on the Heisenberg algebra. The useful deformations
 can be  found for the one  dimensional case in
S.Majid \cite{Majid} and G.Vitiello \cite{Vitiello}, written in terms of annihilation and creation
operators.

In our case the class of possible deformations will be parametrized by a
Bogoljubov operator $\chi$ on the classical phase space $\Gamma$ assuming that
a vacuum is given by the choice of a complex structure ${J}$ on
$\Gamma$. The Bogoljubov operator is characterized by anticommuting with the
complex structure ${J}$ as well as with the symplectic form $\sigma$ on
$\Gamma$ (\ref{anticom1}, \ref{anticom2}):

\begin{align*}
{J}\circ \chi &= - \chi \circ {J} \\
\sigma \circ \chi &= - \chi \circ \sigma
\end{align*}

Our deformations will break the manifest symplectic group symmetry of the
extended Heisenberg algebra $\mathbf{U}$ since $\chi$ is not an invariant under
these symmetries. They will be written in terms of a set of
$\mathbb{R}$-independent eigenvectors $\{{{z}_{i},{J}{z}_{i}}\}$.

The deformed commutation relations are:

\begin{align}
 \left[ \phi({z}_{i}),\pi({z}_{j}) \right]  &= -1{\delta}_{ij}{\left[ 2 H
\right]}_{{\chi}_{i}}&\\
\left[ \phi({z}_{i}), H  \right]   &= 0&\\
\left[ \pi({z}_{i}), H  \right]   &= 0 &\text{ where $\pi({z}_{i}) :=
\phi({J}{z}_{i})$}\\
&&\text{ and ${\left[ x \right]}_{{\chi}_{i}} := {\frac{\sinh{{\chi}_{i} x}}
{\sinh{{\chi}_{i}}}}$}
\end{align}

The deformed comultiplication $ {\Delta}_{\chi}$ is:

\begin{align}
{\Delta}_{\chi} \phi({z}_{i}) &= \phi({z}_{i}) \otimes {e}^{{\chi}_{i}H} +
{e}^{{-\chi}_{i}H} \otimes \phi({z}_{i})\\
{\Delta}_{\chi} \pi({z}_{i}) &= \pi({z}_{i}) \otimes {e}^{{\chi}_{i}H} +
{e}^{{-\chi}_{i}H} \otimes \pi({z}_{i})\\
{\Delta}_{\chi} H &= H \otimes \mathbf{1} +  \mathbf{1} \otimes H
\end{align}

It can be checked by direct calculation that the comultiplication $\Delta$
preserves the commutation relations and that it is coassociative:

\begin{align}
\left[ {\Delta}_{\chi} \phi({z}_{i}), {\Delta}_{\chi} \pi({z}_{i}) \right] &=
-1{\delta}_{ij}{\left[ 2 {\Delta}_{\chi} H \right]}_{{\chi}_{i}}\\
({\Delta}_{\chi}\otimes\mathbf{1})\circ {\Delta}_{\chi} &=
(\mathbf{1}\otimes{\Delta}_{\chi} )\circ{\Delta}_{\chi}
\end{align}

In the Weyl form the deformed commutation relations and the comultiplication
can be written as:

\begin{align}
U({z}_{i}) V({z}_{j}) &= {e}^{-i{\left[ 2 H \right]}_{{\chi}_{i}}} V({z}_{j})
U({z}_{i})\\
{\Delta}_{\chi} U({z}_{i}) &= U({e}^{\chi}{z}_{i}) \otimes
U({e}^{-\chi}{z}_{i})\\
{\Delta}_{\chi} V({z}_{i}) &= V({e}^{\chi}{z}_{i}) \otimes
V({e}^{-\chi}{z}_{i})\\
\intertext{with:}
U({z}_{i}) &:= {e}^{i\phi({z}_{i})} = W({z}_{i})\\
V({z}_{i}) &:= {e}^{i\pi({z}_{i})} = W({J}{z}_{i})
\end{align}

We turn now to the canonical mappings $p$, ${p}_{n}$ of the deformed extended
Heisenberg algebra ${\mathbf{U}}_{\chi}$ and its coproducts
${{\Delta}_{\chi}}^{n-1}{\mathbf{U}}_{\chi}$ onto the Heisenberg algebra
$\mathbf{A}$ and its tensor products ${\mathbf{A}}^{\otimes n}$. It will be
required that $p$, ${p}_{n}$ are algebra homomorphisms and that $H$ as well as
${{\Delta}_{\chi}}^{n-1}H$ are mapped by $p$, ${p}_{n}$into the units
$\mathbf{1}$, ${\mathbf{1}}^{\otimes n}$. In the following the generators $\phi({z}_{i})$, $\pi({z}_{i})$ will be identified with their images
$p(\phi({z}_{i}))$, $p(\pi({z}_{i}))$.

      The map $p$ is fully specified and it is thus tempting to set ${p}_{n} =
{p}^{\otimes n}$, but then ${{\Delta}_{\chi}}^{n-1}H$ would be mapped into $n
\cdot \mathbf{1}$ instead of $\mathbf{1}$. To fix the normalization one has to set:

\begin{align}
{p}_{n}({{\Delta}_{\chi}}^{n-1}\phi({z}_{i})) &= \frac 1 {\sqrt{n}}
{p}^{\otimes n}({{\Delta}_{\chi}}^{n-1}\phi({z}_{i}))\\
{p}_{n}({{\Delta}_{\chi}}^{n-1}\pi({z}_{i})) &= \frac 1 {\sqrt{n}}
{p}^{\otimes n}({{\Delta}_{\chi}}^{n-1}\pi({z}_{i}))
\end{align}

Now ${p}_{n}$ is also fully specified. The important thing now is that the map
${{\Delta}_{\chi}}^{n-1}:{\mathbf{U}}_{\chi}\rightarrow
{{\mathbf{U}}_{\chi}}^{\otimes n}$ factors through the maps $p$, ${p}_{n}$ as
can be checked on the generators. The result is a map

\begin{equation}
{{\Delta}_{\chi}}^{(n-1)}:\mathbf{A}\rightarrow {\mathbf{A}}^{\otimes n}
\end{equation}

which fills in the commutative diagram

\begin{equation}
\begin{CD}
{\mathbf{U}}_{\chi} @>p>> \mathbf{A}\\
@V{{{\Delta}_{\chi}}}^{n-1}VV    @VV{{{\Delta}_{\chi}}^{(n-1)}}V\\
{{\mathbf{U}}_{\chi}}^{\otimes n} @>>{{p}_{n}}> {{\mathbf{A}}^{\otimes n}}
\end{CD}
\end{equation}

The map ${{\Delta}_{\chi}}^{(n-1)}$ is an algebraic homomorphism and allows thus
to take tensor products of representations. Note however, that due to the
necessary normalization ${{\Delta}_{\chi}}^{(n-1)}$ is not a comultiplication.

In the case $n=2$, ${{\Delta}_{\chi}}^{(1)}$ is our canonical doubling. If we
use now this canonical doubling for doubling a vacuum state
${\omega}_{{J}}$ and the corresponding Fock representation, one obtains the state ${\omega}_{\chi}$:

\begin{align*}
{\omega}_{\chi}(U({z}_{i}))&\equiv{\omega}_{\chi}(W({z}_{i})) =&\\
&=({\omega}_{ J}\otimes {\omega}_{ J})\circ
{{\Delta}_{\chi}}^{(1)}(U({z}_{i})) =&\\
&=({\omega}_{ J}\otimes {\omega}_{ J})(U({e}^{\chi}\frac
{{z}_{i}} {\sqrt{2}} )\otimes U({e}^{\chi}\frac {{z}_{i}} {\sqrt{2}} )) =&\\
&={e}^{-\frac 1 4 (\frac 1 2 {z}_{i}\circ {e}^{\chi}\circ{J}\circ
\sigma\circ {e}^{\chi}\circ {z}_{i})} {e}^{-\frac 1 4 (\frac 1 2 {z}_{i}\circ
{e}^{-\chi}\circ{J}\circ \sigma\circ {e}^{-\chi}\circ {z}_{i})} =&\\
&={e}^{-\frac 1 4 ({z}_{i}\circ \frac {({e}^{2\chi}+{e}^{-2\chi})} 2
\circ{J}\circ \sigma\circ {z}_{i})} =&\\
&={e}^{-\frac 1 4 ({z}_{i}\circ \cosh{2\chi} \circ{J}\circ \sigma\circ
{z}_{i})} &\\
\intertext{and similarly:}
{\omega}_{\chi}(V({z}_{i}))&\equiv {\omega}_{\chi}(W({J}{z}_{i}))=&\\
&={e}^{-\frac 1 4 ({J}{z}_{i}\circ \cosh{2\chi} \circ{J}\circ
\sigma\circ {J}{z}_{i})}&
\end{align*}

By extension from the generators one gets:

\begin{equation}\label{resultstate}
{\omega}_{\chi}(W({z}_{i})) = {e}^{-\frac 1 4 ({z}_{i}\circ \cosh{2\chi}
\circ{J}\circ \sigma\circ {z}_{i})}
\end{equation}

But this is just the quasifree state (\ref{quasifree}).

\section{An Example: The Harmonic oscillator}\label{4}

In  the  special  case  of  a  1-dimensional  harmonic  oscillator
some  particular  simplifications  occur.  In  its 2-dimensional
phase space  $\Gamma \cong {\mathbb{R }  }^{2 } $ there  exists a
basis  in  which  the  symplectic  form  $\sigma  $,  the complex
structure ${J} $ and the given Bogoljubov operator $\chi $ take the
form:

\begin{align}
{\sigma}_{ab} &= \begin{pmatrix}
0 & 1 \\
-1 & 0
\end{pmatrix}\\
{{ J}^{a}}_{b} &= \begin{pmatrix}
0 & 1 \\
-1 & 0
\end{pmatrix}\\
{{\chi}^{a}}_{b} &= \begin{pmatrix}
\chi & 0 \\
0 & -\chi
\end{pmatrix}
\end{align}
This  basis is  unique  and  can  be  given  also by geometrical
considerations unless  $\chi = 0 $.  Using ${J} $ as  the imaginary
unit  one can  now identify  the phase  space $\Gamma  $ with the
complex numbers:

\begin{equation}
\Gamma \cong \mathbb{C}
\end{equation}

The vacuum state ${\omega }_{ J}$ and the state ${\omega }_{\chi }$
obtained in (\ref{resultstate}) are then given by:

\begin{align}
{\omega}_{ J}(W(z)) &= {e }^{-\frac {1 } {4 } {\|z\|}^{2}}\\
{\omega}_{\chi}(W(z))  &= {e  }^{-\frac  {1  } {4  } \cosh{2\chi}
{\|z\|}^{2}}\label{chistate}
\end{align}

The   Hamiltonians   compatible   with   the   complex  structure
${J}$ (i.e.
those which have a diagonalization (\ref{Hamiltonian}) giving the
fixed ${J}$) are determined by the matrix

\begin{equation}
{ H}_{ab} = E \beta \begin{pmatrix}
1 & 0 \\
0 & 1
\end{pmatrix}
\end{equation}

The corresponding thermal states are

\begin{equation}
{\omega }_{\beta }(W(z)) ={e}^{-\frac  {1} {4} \coth{\frac {\beta
E}{2}} {\|z\|}^{2}} \label{betastate}
\end{equation}

Fixing $E$ as the energy of the harmonic oscillator and comparing
(\ref{chistate}) and (\ref{betastate})  the 
states  ${\omega}_{\chi}$ are identified with  the thermal states, ${\omega
}_{\beta }$, of the Hamiltonian at inverse temperature $\beta$,
obtaining the relation

\begin{equation}
\coth{\frac {\beta E}{2}}= \cosh{2 \chi}
\end{equation}

between  the inverse  temperature  $\beta  $ and  the deformation
parameter  $\chi $.  So, in  the end,  the deformation  parameter
$\chi $  has a nice interpretation  as a function of  the inverse
temperature.

\section{Conclusion}\label{5}
      It is shown that there is a class of deformations of the extended
Heisenberg bialgebra ${\mathbf{U}}_{\chi}$ that provide canonical doublings by mapping down their
comultiplications  on the Heisenberg algebra $\mathbf{A}$.
These doublings give directly, without any redefinitions, all representations
arising from quasifree states by the GNS-construction. In particular the
doublings give all thermal representations for free Hamiltonians.
      Our construction works clearly for any finite dimensional system and on
the level of calculus also for infinite dimensional systems. Functional
analytic discussions for the infinite dimensional case are omitted. Note,
however, that at no point  unitary equivalence of
representations is used and that all Bogoljubov transformations are given by
symplectomorphisms on the classical phase space. Thus no problem is expected in extending our considerations to the infinite dimensional case.

      It would be useful to know if there are other deformations other, than the ones used here, of the
extended Heisenberg algebra $\mathbf{U}$. If not,
then the construction becomes entirely canonical, since the class of
deformations appears to be the only choice one could make in the construction.
If there are other deformations then it would be interesting to see the
interpretation of the induced doublings on the Heisenberg algebra $\mathbf{A}$
arising from them.

      In any  case the present  results show that  the bialgebra
structure is a logical way to approach Thermal Field Theory by
providing both the correct results and a mathematically satisfactory general
structur.

\section{Acknowledgements}
Two of us (F.C.K. and A.E.S.) would like to thank G.Vitiello and M.Tuite for
useful discussions and one of us (T.K.) would like to thank T. Brzezi\'nski and
P.Krtou\v{s} for having explained to him several steps used here and to D.N.Page
for useful discussions and support.

The research is supported in part by Natural Sciences and Engineering Research Council of Canada and by CAPES (a Brazilian Government Agency for Research).


\begin{thebibliography}{88}
  \bibitem{Umezawa}Y.Takahashi and H.Umezawa, Collective Phenomena {\bf 2},  55  (1995);\\
H.Umezawa, H.Matsumoto and M.Tachiki, {\em Thermo Field Dynamics and Condensed States}
  (North Holland, Amsterdam, 1982);\\
H.Umezawa, {\em Advanced Field theory:Micro, Macro and Thermal Physics} (AIP,
  New York, 1993) \\
and references therein.

  \bibitem{Ojima}I.Ojima, Ann. Phys. {\bf 137},  1  (1981).

  \bibitem{Madore}J.Madore, {\em An Introduction to Noncommutative Geometry and its Physical
  Applications} (London Mathematical Society, Cambridge University Press,
  Cambridge, 1995).

  \bibitem{Khanna_Santana}A.E.Santana and F.C.Khanna, Phys. Lett. {\bf A 203},  68  (1995).

 \bibitem{Song-Ding-An}H.~S.~Song, S.~X.~Ding and I. An, J. Phys. A: Math. Gen. {\bf 26},  5197  (1993).

\bibitem{Iorio-Vitiello}A.Iorio and G.Vitiello, Mod.Phys.Lett. {\bf B8}, 269 (1994).

\bibitem{Celeghini1}E. Celeghini, S. de Martino, S. de Siena, G. Vitiello and M. Rasetti, Mod. Phys. Lett.  {\bf B7}, 1321 (1993)

\bibitem{Celeghini2}E. Celeghini, S. de Martino, S. de Siena, M. Rasetti  and G. Vitiello, Ann. Phys. {\bf 241}, 50 (1995)

   \bibitem{Bratteli-Robinson1}O.Bratteli and D.W.Robinson, {\em Operator Algebras and Quantum Statistical
  Mechanics} (Springer-Verlag, New York, 1979), Vol.~I.

  \bibitem{Takesaki}M.Takesaki, {\em Tomita's theory of modular Hilbert algebras and its
  applications}, Vol.~128 of {\em Lecture Notes in Mathematics}
  (Springer-Verlag, Berlin-Heidelberg-New York, 1970).

  \bibitem{Bratteli-Robinson2}O.Bratteli and D.W.Robinson, {\em Operator Algebras and Quantum Statistical
  Mechanics} (Springer-Verlag, New York, 1981), Vol.~II.

  \bibitem{Krtous}P.Krtou\v{s}:gr-qc/9507023, private communication

  \bibitem{Majid}S.Majid: {\em Foundations of Quantum Group Theory}, (Cambridge
University Press, 1995)

  \bibitem{Vitiello}G.Vitiello: The algebraic structure of Thermo Field
Dynamics and the $q$-deformation of the Weyl-Heisenberg algebra, 1995, in  the {\em Proceedings of the IV. Workshop on Thermal Theories and Their Applications}, eds. Y.x.Gui, F.C.Khanna and Z.B.Su (World Scientific, Singapore, 1996), p.13.

\end{thebibliography}
\end{document}